\documentclass[aps,prd,preprintnumbers,superscriptaddress,nofootinbib,showpacs,twocolumn]{revtex4}%
\usepackage[dvipdfmx]{graphicx}
\usepackage{bm,latexsym,amsmath,amssymb,amsfonts,mathrsfs}
\usepackage{color}
\input{colordvi.tex}
\usepackage[dvipdfmx]{hyperref}
\usepackage{url}
\hypersetup{
    colorlinks=true,
    citecolor=cyan,
}
\newcommand*{\D}{{\rm d}}
\newcommand*{\mpl}{M_{\rm Pl}}
\newcommand*{\tilh}{\widetilde{h}_0}
\begin{document}

\title{Reheating and Primordial Gravitational Waves in Generalized Galilean Genesis}

	%Matter Creation in Generalized Galilean Genesis

\author{Sakine~Nishi}
\email[Email: ]{sakine"at"rikkyo.ac.jp}
\affiliation{Department of Physics, Rikkyo University, Toshima, Tokyo 171-8501, Japan
}

\author{Tsutomu~Kobayashi}
\email[Email: ]{tsutomu"at"rikkyo.ac.jp}
\affiliation{Department of Physics, Rikkyo University, Toshima, Tokyo 171-8501, Japan
}

\begin{abstract}
Galilean genesis is an alternative to inflation, in which the universe starts
expanding from Minkowski with
the stable violation of the null energy condition. 
In this paper,
we discuss how the early universe is reheated through the gravitational particle production
at the transition from the genesis phase to the subsequent phase where the kinetic energy
of the scalar field is dominant.
We then study the consequences of gravitational reheating after Galilean genesis
on the spectrum of primordial gravitational waves.
The resultant spectrum is strongly blue, and at high frequencies $\Omega_{\rm gw}\propto f^3$
in terms of the energy density per unit logarithmic frequency.
Though this cannot be detected in existing detectors,
the amplitude can be as large as $\Omega_{\rm gw}\sim 10^{-12}$ at $f\sim 100\,$MHz,
providing a future test of the genesis scenario.
The analysis is performed within the framework of generalized Galilean genesis
based on the Horndeski theory, which enables us to derive generic formulas.
\end{abstract}

\preprint{RUP-16-3}
\maketitle
%%%%%%%%%%%%%%%%%%%%%%%%%%%%%%%%%%%%%%%%%%%%%%%%%%%%%%%%%%%%%%%%%%%%%%

%--- Introduction ---%
\section{Introduction}

%Inflation
Inflation~\cite{Guth:1980zm,Sato:1980yn} is now the standard model of the early universe,
solving the problems that Big Bang cosmology faces and
explaining the origin of large scale structure in the way consistent with observations.
However, the inflationary universe still suffers from
the problem of initial singularity~\cite{Borde:1993xh} as long as the null energy condition (NEC) is satisfied.
To circumvent the singularity, alternative scenarios
such as the bouncing models~\cite{Brandenberger:2009jq,Brandenberger:2010dk,Brandenberger:2011gk,deHaro:2015wda}
have been proposed so far, but
a majority of the alternative models are unstable
due to the violation of the NEC.
Recently, an interesting class of scalar field theories
called the Galileon has been constructed~\cite{Nicolis:2008in}
and various aspects of the Galileon theory have been explored extensively in the literature.
One of the most intriguing nature of this theory is that
the null energy condition can be violated stably\footnote{See Ref.~\cite{Sawicki:2012pz} for
a subtle point beyond linear perturbations.}.
Thus, the Galileon opens up a new possibility of
stable, singularity-free models alternative to inflation~\cite{Qiu:2011cy,Easson:2011zy,Cai:2012va,Cai:2013vm,Osipov:2013ssa,Qiu:2013eoa},
as well as stable NEC violating models of dark energy~\cite{Deffayet:2010qz}
and inflation~\cite{Kobayashi:2010cm}.

%Genesis/%Reheating
In this paper, we consider an initial phase of NEC violating quasi-Minkowski expansion
driven by the Galileon field as an alternative to inflation.
The scenario is called the Galilean genesis~\cite{Creminelli:2010ba},
and different versions of genesis models can be found in
Refs.~\cite{Liu:2011ns,Creminelli:2012my,Hinterbichler:2012fr,Hinterbichler:2012yn}.
The purpose of the present paper is to
address how the genesis phase is connected to the subsequent hot universe through the reheating stage.
In the case of inflation, reheating usually proceeds with coherent oscillation of
the scalar field at the minimum of the potential and its decay.
However, the scalar field that drives Galilean genesis does not have the potential.
The situation here is similar to that in kinetically driven inflation models
such as k-~\cite{ArmendarizPicon:1999rj} and G-inflation~\cite{Kobayashi:2010cm}.
We therefore consider reheating through gravitational particle production~\cite{Ford:1986sy}.
Reheating after Galilean genesis has also been discussed in Ref.~\cite{LevasseurPerreault:2011mw}.
See also Refs.~\cite{Easson:2013bda,Pirtskhalava:2014esa,Liu:2014tda,Kobayashi:2015gga}
for the other aspects of genesis models.

The exit from k- and G-inflation is accompanied by the kination phase,
{\em i.e.,} the phase where the kinetic energy of the scalar field is dominant.
Similarly, the kination phase can also be incorporated at the end of Galilean genesis
as we will do in this paper.
The primordial gravitational waves that re-enter the horizon during this kination era
have a blue spectrum, giving enhanced amplitudes at high frequencies~\cite{Tashiro:2003qp}.
In addition to this, the primordial spectrum of gravitational waves
is expected to be blue due to NEC violating quasi-Minkowski expansion
in the earliest stage of the universe. The combined effect will therefore give rise to
strongly blue gravitational waves.
In the second part of this paper we evaluate the spectrum of gravitational waves
generated from Galilean genesis and explore the possibility of testing this alternative scenario
{\em e.g.,} with the advanced LIGO detector.

In the previous paper~\cite{Nishi:2015pta}, we have developed
a general framework unifying the original genesis model~\cite{Creminelli:2010ba}
and its extensions~\cite{Liu:2011ns,Creminelli:2012my,Hinterbichler:2012fr,Hinterbichler:2012yn},
using which we have derived the stability conditions and
primordial power spectra of scalar and tensor perturbations.
The framework is based on the Horndeski theory~\cite{Horndeski,Deffayet:2011gz,Kobayashi:2011nu},
the most general second-order scalar-tensor theory in four dimensions.
We use this general framework also in this paper.

%\cite{Pirtskhalava:2014esa,Liu:2014tda,Kobayashi:2015gga}

%plan of this paper
The plan of this paper is as follows. 
In the next section, we review the Galilean genesis solution and its generalization.
In Sec.~III, we investigate the creation of massless scalar particles
after Galilean genesis
and compute the reheating temperature.
Using the result of Sec.~III, we then evaluate the power spectrum of gravitational waves in Sec.~IV.
We present some concrete examples and discuss the detectability of
the primordial gravitational waves in Sec.~V.
% in the cases of 
%$\alpha =1$ as a original genesis model and $\alpha =2$ which gives the scale invariant powerspectrum of scalar perturbation,
%and compare to the sensitivity of advanced LIGO.
Finally, we conclude in Sec. VI.

%--- Generalized genesis solutions ---%
\section{Generalized genesis solutions}
Galilean genesis was originally proposed by Creminelli {\em et al.}~\cite{Creminelli:2010ba}, 
and several versions of genesis models have been considered so far~\cite{Liu:2011ns,Creminelli:2012my,Hinterbichler:2012fr,Hinterbichler:2012yn}. 
In the Galilean genesis scenario, the universe starts expanding
from Minkowski spacetime with a stable violation of the null energy condition.
In the previous work~\cite{Nishi:2015pta}, we have
constructed a unified description of those genesis models
based on the Horndeski theory and named the scenario {\em generalized Galilean genesis}.
The Lagrangian of the Horndeski theory is given by~\cite{Horndeski,Deffayet:2011gz,Kobayashi:2011nu}
\begin{align}
{\cal L}&=G_2(\phi, X)-G_3(\phi, X)\Box\phi+G_4(\phi, X)R
\notag \\ & \quad
+G_{4,X}\left[(\Box\phi)^2-(\nabla_\mu\nabla_\nu\phi)^2\right]
+G_5(\phi, X)G^{\mu\nu}\nabla_\mu\nabla_\nu\phi
\notag \\ & \quad
-\frac{1}{6}G_{5,X}\left[(\Box\phi)^3
-3\Box\phi(\nabla_\mu\nabla_\nu\phi)^2
+2(\nabla_\mu\nabla_\nu\phi)^3\right],
\label{Hor_L}
\end{align}
where $X:=-(1/2)g^{\mu\nu}\partial_\mu\phi\partial_\nu\phi$.
This Lagrangian gives the most general second-order field equations
for the metric and the scalar field.
Generalized galilean genesis
is obtained by taking the four functions in the Horndeski Lagrangian~(\ref{Hor_L}) as
\begin{eqnarray}
&&
G_2=e^{2(\alpha+1)\lambda\phi}g_2(Y),
\quad
G_3=e^{2\alpha\lambda\phi} g_3(Y),
\nonumber\\&&
G_4=\frac{M^2_{\rm Pl}}{2}+e^{2\alpha\lambda\phi}g_4(Y),
\quad
G_5=e^{-2\lambda\phi}g_5(Y),\label{gen:Lag}
\end{eqnarray}
where $\alpha\;(>0)$ and $\lambda$ are constant parameters, $\mpl$ is the Planck mass,
and each $g_i$ ($i=2,3,4,5$) is an arbitrary function of
\begin{eqnarray}
Y:= e^{-2 \lambda\phi} X.
\end{eqnarray}

The above Lagrangian admits the generalized genesis solution
for the scalar field and the cosmic scale factor $a(t)$,
\begin{eqnarray}
e^{\lambda\phi}&\simeq& \frac{1}{\lambda\sqrt{2Y_0}}\frac{1}{(-t)}, \label{gen:back_e} \\
H&\simeq&\frac{h_0}{(-t)^{2\alpha+1}}, \label{gen:back_H} \\
a &\simeq& a_G\left[1+\frac{1}{2\alpha}\frac{h_0}{(-t)^{2\alpha}}\right],\label{gen:back_a}
\end{eqnarray}
where $Y_0$ and $h_0$ are constants determined in the way explained below and $H:=\D \ln a/\D t$
is the Hubble parameter.
The proper time coordinate $t$ lies in the range $(-\infty, 0)$,
and the scale factor approaches to a constant value $a_G$ as $t\to -\infty$.
Thus, this solution describes a universe that starts expanding from Minkowski in the asymptotic past.
The parameter $\alpha$ controls the expansion rate.
It has been shown in~\cite{Nishi:2015pta} that
the generalized genesis solution is an attractor for a wide range of initial conditions.
Note that one may take $a_G=1$, though we do not do so in this paper because
we would rather normalize the scale factor to be unity at the present epoch.
Note also that the above solution is the approximate one valid for
\begin{eqnarray}
\frac{h_0}{(-t)^{2\alpha}}\ll 1.\label{gen-an-cond}
\end{eqnarray}
The background evolution begins to deviate from the above solution when $h_0/(-t)^{2\alpha}\sim 1$,
and one need to invoke a numerical calculation for a later epoch.
To make the calculations tractable analytically, in this paper we
impose the technical assumption that the genesis phase
terminates before Eq.~(\ref{gen-an-cond}) is violated.
Denoting the Hubble parameter at the end of genesis by $H_\ast$,
the condition~(\ref{gen-an-cond}) can be written as
\begin{eqnarray}
H_\ast\ll h_0^{-1/2\alpha},\label{gen-an-cond2}
\end{eqnarray}
or, equivalently,
\begin{eqnarray}
H_\ast(-t_\ast)\ll 1,
\end{eqnarray}
where $t_\ast$ is the proper time at the end of the genesis phase.
This inequality defines the range of validity of our computation.

The constants $Y_0$ and $h_0$ are consistently determined
from
the Friedmann and evolution equations,
\begin{eqnarray}
\hat\rho(Y_0)&\simeq& 0,\label{gen:eq1} \\
-2{\cal G}(Y_0)\dot H&\simeq&
e^{2(\alpha +1)\lambda\phi}\hat p(Y_0),\label{gen:eq2}
\end{eqnarray}
where
\begin{eqnarray}
\hat\rho(Y)&:=&2Y g_2'-g_2-4\lambda Y\left(\alpha g_3-Yg_3'\right),
\\
\hat p(Y)&:=&g_2-4\alpha \lambda Yg_3 
\nonumber\\&&+8(2\alpha +1)\lambda^2 Y(\alpha g_4-Yg_4') ,
\\
{\cal G}(Y)&:=&M^2_{\rm Pl}-4\lambda Y\left(g_5+Yg_5'\right).
\end{eqnarray}
The stability condition for tensor perturbations requires ${\cal G}(Y_0)>0$,
so that the NEC violating background can be obtained if $\hat p(Y_0)<0$.
One sees
from~(\ref{gen:eq2}) that
\begin{eqnarray}
h_0= -\frac{1}{2(2\alpha+1)(2\lambda ^2Y_0)^{\alpha +1}}\frac{\hat p(Y_0)}{{\cal G}(Y_0)}.\label{h0determined}
\end{eqnarray}
Since $h_0$ has the dimension of [mass]$^{-2\alpha}$,
it is convenient to introduce the dimensionless quantity $\tilh$ defined as
\begin{align}
\tilh :=\mpl^{2\alpha}h_0.\label{Bmu}
\end{align}

% and
%${\cal G}\sim \mpl^2$, it is convenient to write
%\begin{eqnarray}
%h_0=B\mpl^{-2} \mu^{-2\alpha +2},\label{Bmu}
%\end{eqnarray}
%where $B$ is a dimensionless constant and $\mu$ has the dimension of mass.
%The values of $B$ and $\mu$ are fixed once we specify a concrete
%Lagrangian for the genesis scinario.
%Note that $B$ is not necessarily of ${\cal O}(1)$
%because an underlying Lagrangian may contain two different mass scales (other than $\mpl$)
%and so in some cases it is not appropriate to write $h_0$ solely
%in terms of a single mass scale $\mu$.
%Using $B$ and $\mu$ one can rewrite Eq.~(\ref{gen-an-cond2}) as
%\begin{eqnarray}
%\frac{H_\ast}{\mpl}\ll B^{-1/2\alpha}\left(\frac{\mu}{\mpl}\right)^{(\alpha-1)/\alpha}
%\label{techcons}.
%\end{eqnarray}

Let us summarize here the stability conditions derived in Ref.~\cite{Nishi:2015pta}.
For the stability of tensor perturbations, we have to impose
\begin{eqnarray}
{\cal G}(Y_0)>0,\quad
M^2_{\rm Pl}+4\lambda Y_0 g_5(Y_0)>0.\label{co4}
\end{eqnarray}
Scalar perturbations are stable if
\begin{eqnarray}
\hat\rho'(Y_0)>0,
\quad
\xi'(Y_0)<0,\label{co6}
\end{eqnarray}
where $\xi$ is defined as
\begin{eqnarray}
\xi(Y):=-\frac{Y{\cal G}(Y)}{\hat p(Y)}.
\end{eqnarray}

The universe must eventually be connected to a radiation dominated phase.
In standard inflationary cosmology, the inflaton field oscillates about the minimum of
the potential and reheat the universe. The radiation dominated universe follows after this reheating stage.
However, the Lagrangian for generalized Galilean genesis does not have the potential term
and hence conventional reheating would not be suitable for the genesis scenario.
This is also the case for k-inflation~\cite{ArmendarizPicon:1999rj}.
In this paper, we therefore explore the possibility of gravitational reheating
after the genesis phase, {\em i.e.}, reheating via gravitational particle production
at the transition to the intermediate phase where the kinetic energy of the scalar field
is dominant (the kination phase).
See~\cite{LevasseurPerreault:2011mw} for preheating after Galilean genesis
via a direct coupling between the scalar field and the other fields.
Note in passing that
we have another possibility that
the genesis phase is followed by inflation~\cite{Pirtskhalava:2014esa,Liu:2014tda,Kobayashi:2015gga}.
The inflationary stage is then naturally described by k- or G-inflation~\cite{Kobayashi:2010cm},
so that also in this case we employ gravitational reheating.
In this paper, we do not consider this latter possibility and
focus on the scenario in which the genesis phase is followed by
the intermediate kination phase and then by the radiation dominated phase.

%--- Gravitational particle production ---%
\section{Gravitational particle production after generalized Galilean genesis}
As argued at the end of the previous section, we consider
gravitational particle production after the genesis phase.
To realize the background evolution suitable to this reheating mechanism,
we assume that the Lagrangian of the form~(\ref{gen:Lag})
is only valid until $\phi$ (which is increasing during the genesis phase)
reaches some value, $\phi_\ast$, and for $\phi>\phi_\ast$ the Lagrangian is
such that the kinetic energy of $\phi$ decreases quickly
to make only the standard kinetic term $X$ relevant in the Lagrangian:
${\cal L}\simeq (\mpl^2/2)R+X$.
The genesis phase thus terminates and is followed by the kination phase where
$3\mpl^2H^2=\rho_\phi\propto a^{-6}$.\footnote{%
The necessary requirement here is that
the energy density of $\phi$ dilutes more rapidly than that of radiation.
We assume the kination phase just for simplicity.}
The basic scenario here is essentially the same as
those at the end of k- and G-inflation~\cite{ArmendarizPicon:1999rj,Kobayashi:2010cm}
described in some detail in \cite{Kunimitsu:2012xx}.
We do not provide a concrete model realizing this because
it is reasonable to presume that this is indeed possible
by using the functional degrees of freedom in the Horndeski theory,\footnote{%
We assume the particular form of the
Lagrangian~(\ref{gen:Lag}) only during the genesis phase.
At the very end of genesis, the Lagrangian may be away from the form~(\ref{gen:Lag}).}
and also because particle creation is only sensitive to
the evolution of the scale factor but not to an underlying concrete model of generalized Galilean genesis.
Note that in many cases it is likely that perturbations become unstable at the transition between the two phases.
Even if this occurs we can solve the issue by using the idea in Ref.~\cite{Kobayashi:2015gga},
though one must go beyond the Horndeski theory to do so.

Let us consider the creation of massless, minimally coupled scalar particles
at the transition from the genesis phase to kination.
The created particle is denoted by $\chi$, whose the Lagrangian is given by
\begin{eqnarray}
{\cal L}_\chi=-\frac{1}{2}g^{\mu\nu}\partial_\mu\chi \partial_\nu\chi.
\end{eqnarray}
Its Fourier component, $\chi_k$, obeys
\begin{align}
\frac{1}{a}\left(a\chi_k\right)''+\left(k^2-\frac{a''}{a}\right)\chi_k=0,
\end{align}
where a dash stands for differentiation with respect to the conformal time $\eta$.
Since $a\simeq a_G=\;$const in the genesis phase, we have $\eta\simeq t/a_G$ and hence $\eta=-\infty$
corresponds to the asymptotic past.
We write the solution using the Bogoliubov coefficients as
\begin{eqnarray}
a\chi_k = \frac{\alpha_k(\eta)}{\sqrt{2k}}e^{-ik\eta}
+\frac{\beta_k(\eta)}{\sqrt{2k}}e^{ ik\eta},
\end{eqnarray}
and impose the boundary conditions that $\alpha_k\to 1$ and $\beta_k\to 0$ as $\eta\to -\infty$.
For $k^2 \gg |a''/a|$, we have~\cite{BDqunat,Parker:1969au,Zeldovich:1971mw,Zeldovich1977}
\begin{eqnarray}
\beta_k(\eta) = -\frac{i}{2k}\int_{-\infty}^\eta e^{-2iks}\frac{a'' }{a }\D s.
\end{eqnarray}
The energy density of the created particles can be computed as
\begin{eqnarray}
\rho_\chi = \frac{1}{2\pi^2 a^4}\int_0^\infty k^3\left|\beta_k(\infty)\right|^2 \D k,
\end{eqnarray}
and this can be recasted to
\begin{align}
\rho_\chi &=-\frac{1}{128\pi^2a^4} 
\int^\infty_{-\infty}\D \eta_1\int^\infty_{-\infty}\D \eta_2 \ln(m|\eta_1-\eta_2|)
 \notag \\
& \quad\times
V'(\eta_1)V'(\eta_2),
\label{rho_int}
\end{align}
where
\begin{eqnarray}
V(\eta)=\frac{f''f-(f')^2/2}{f^2},
\end{eqnarray}
with $f(\eta):=a^2(\eta)$
and we inserted some arbitrary mass scale $m$ in $\ln$
for a dimensional reason, though $\rho_\chi$ will be dependent only logarithmically on $m$.

In the genesis phase,
we have
\begin{eqnarray}
f(\eta)\simeq a_G^2\left[1+\frac{h_0}{\alpha}\frac{1}{(-a_G\eta)^{2\alpha}}\right]
\quad(\eta < \eta_\ast),
\end{eqnarray}
with $\eta_\ast:=t_\ast/a_G$, while in the subsequent kination phase
$a\propto (\eta+{\rm const})^{1/2}$ and hence $f$ is of the form
\begin{eqnarray}
f(\eta)=c_0\frac{\eta}{-\eta_\ast} + c_1,
\end{eqnarray}
where $c_0$ and $c_1$ are to be determined by requiring that
$f$ and $f'$ are continuous at $\eta=\eta_\ast$.
Under this ``sudden transition'' approximation,
$f''$ is discontinuous and consequently $V$ is also discontinuous at $\eta=\eta_\ast$.
Due to this discontinuity, the integral~(\ref{rho_int}) diverges,
which is unphysical. To avoid divergence, we insert a short stage
having a time scale $\Delta \eta$ in between the genesis and kination phases,
and join the two phases smoothly. 
The idea here is also employed in connecting inflation
to the radiation/kination phase smoothly~\cite{Ford:1986sy,Kunimitsu:2012xx}.

Thus, we make the ansatz
\begin{eqnarray}
f(\eta)=\left \{
\begin{array}{ll}
b_0+b_1\eta+b_2\eta^2+b_3\eta^3 & (\eta_{*} < \eta < \eta_\ast+\Delta \eta)\\
c_0\eta/(-\eta_\ast)+c_1  & (\eta_\ast+\Delta \eta<\eta)
\end{array}
\right.,
\end{eqnarray}
after the end of genesis,
and determine the six coefficients by requiring
that $f$, $f'$, and $f''$ are continuous at $\eta=\eta_\ast$ and $\eta=\eta_\ast+\Delta \eta$.
This allows us to obtain continuous $V$ and hence finite $\rho_\chi$.
For the current purpose we do not have to write
$b_0$, $b_1$, $b_2$, and $b_3$ explicitly, but we only need
\begin{align}
c_0&= \frac{2a_G^{2}h_0}{ (-t_\ast)^{2\alpha}}
\left[1+{\cal O}\left(\frac{\Delta\eta}{-\eta_\ast}\right)\right],
\\
c_1&=a_G^2\left\{1+\frac{(2\alpha+1)h_0}{\alpha(-t_\ast)^{2\alpha}}
\left[1+{\cal O}\left(\frac{\Delta\eta}{-\eta_\ast}\right)\right]\right\}.
\end{align}
Now we have
\begin{eqnarray}
V(\eta)\simeq 2 a_G^2 (2\alpha+1)h_0 \left(- a_G \eta\right)^{-2(\alpha+1)}
\quad (\eta < \eta_\ast),
\end{eqnarray}
and $|V|\ll V(\eta_\ast)$ for $\eta>\eta_\ast + \Delta\eta$.
To evaluate the integral~(\ref{rho_int}),
it is sufficient to approximate $V$
as a straight line for $\eta_\ast<\eta<\eta_\ast+\Delta \eta$.
Thus,
\begin{align}
V&\simeq a_G^2\frac{2(2\alpha +1)h_0}{(-t_\ast)^{2(\alpha +1)}\Delta \eta}(-\eta+\eta_\ast+\Delta\eta)
\notag\\
&\qquad(\eta_\ast<\eta<\eta_\ast+\Delta \eta).
\end{align}

The main contribution to the integral~(\ref{rho_int}) comes from
the region $\eta_\ast<\eta<\eta_\ast+\Delta\eta$
where $V'$ gets much larger than in the genesis and kination phases.
Performing the integral in this domain, we obtain
\begin{align}
\rho_\chi=\frac{(2\alpha+1)^2}{32\pi^2}\ln\left(\frac{1}{a_GH_\ast\Delta\eta}\right)
\frac{h_0^2}{(-t_\ast)^{4(\alpha+1)}}\left(\frac{a_G}{a}\right)^4.\label{rho_rad_1}
\end{align}
A logarithmic divergence is now manifest in the sudden transition limit, $\Delta\eta\to 0$.
In the case of the smooth transition, however, we may take $\ln(1/a_GH_\ast\Delta\eta)={\cal O}(1)$,
leading to a finite energy density of relativistic particles.
To make Eq.~(\ref{rho_rad_1}) more suggestive, we
rewrite the equation in different ways by
using $-t_\ast = h_0^{1/(2\alpha+1)}H_\ast^{-1/(2\alpha +1)}$
and Eq.~(\ref{Bmu}). First, we have
\begin{align}
\rho_\chi&=\frac{A}{32\pi^2}\tilh^{-\frac{2}{2\alpha +1}}
\left(\frac{H_\ast}{\mpl}\right)^{-\frac{4\alpha}{2\alpha +1}}
H_\ast^4\left(\frac{a_G}{a}\right)^4,
\end{align}
where
\begin{eqnarray}
A:=(2\alpha+1)^2\ln\left(\frac{1}{a_GH_\ast\Delta\eta}\right)
\end{eqnarray}
is a number of ${\cal O}(1)$.
This result is in contrast to that in the inflationary scenario, where
the energy density of the created particles is simply given by $\rho_\chi\sim H_{\rm inf}^4/a^4$,
with $H_{\rm inf}$ being the inflationary energy scale.
We may also write $\rho_\chi$ as
\begin{eqnarray}
\rho_\chi=\frac{A}{128\pi^2\alpha^2}\frac{H_\ast^4}{\delta_\ast^2}\left(\frac{a_G}{a}\right)^4,
\end{eqnarray}
where $\delta_\ast$ is defined as
\begin{eqnarray}
\delta_\ast:=\frac{a(\eta_\ast)}{a_G}-1\;(\ll 1).
\end{eqnarray}
This second expression shows that
actual $\rho_\chi$ is much larger than
the naive estimate deduced from
the case of inflation, $\rho_\chi\sim H_\ast^4(a_G/a)^4$.

The reheating temperature $T_R$ is determined from
$\rho_\chi = \rho_\phi$, where $\rho_\phi$
is the energy density of the scalar field after the end of genesis,
\begin{eqnarray}
\rho_\phi= 3\mpl^2H_\ast^2\left(\frac{a_G}{a}\right)^6.
\end{eqnarray}
The scale factor at the time when $\rho_\chi=\rho_\phi$ occurs, $a_R$, is given by
\begin{align}
\frac{a_R}{a_G}=\sqrt{\frac{96\pi^2}{A}}
\left(\frac{H_\ast}{\mpl}\right)^{-\frac{1}{2\alpha+1}}\tilh^{\frac{1}{2\alpha+1}}.\label{aragratio}
\end{align}
Since $a_R>a_G$, we have the condition
\begin{eqnarray}
\frac{H_\ast}{\mpl}<\left(\frac{96\pi^2}{A}\right)^{(2\alpha +1)/2}
\tilh.
\end{eqnarray}
Equating the radiation energy density at $a=a_R$ to $(\pi^2g_\ast/30)T_R^4$,
where $g_\ast$ is the effective number of relativistic species of particles,
we obtain
\begin{align}
\frac{T_R}{\mpl}
&=\left(\frac{30}{\pi^2g_\ast}\right)^{1/4}
\frac{A^{3/4}}{\sqrt{3}(32\pi^2)^{3/4}}
\tilh^{-\frac{3}{2(2\alpha +1)}}
\left(\frac{H_\ast}{\mpl}\right)^{\frac{\alpha+2}{2\alpha+1}}.
\label{trh1}
\end{align}
This result is again in contrast to that in the inflationary scenario,
where $T_R\sim H_{\rm inf}^2/\mpl$.
One can also write $T_R$ as
\begin{align}
T_R&=\left(\frac{30}{\pi^2g_\ast}\right)^{1/4}
\frac{A^{3/4}}{\sqrt{3}(32\pi^2)^{3/4}(2\alpha)^{3/2}}
\frac{H_\ast^2}{\mpl \delta_\ast^{3/2}}.\label{trh2}
\end{align}
From this one can clearly see that
the reheating temperature is much higher than
the naive estimate deduced from inflation, $T_R\sim H_\ast^2/\mpl$.
In other words, for a fixed reheating temperature,
the Hubble parameter at the end of genesis is smaller than the
corresponding inflationary value.

%--- Gravitational waves ---%
\section{The spectrum of primordial gravitational waves}
Having obtained the background evolution in the previous section, 
let us discuss the spectrum of primordial gravitational waves
from generalized Galilean genesis. As expected from the
null energy condition violating nature of the scenario,
generated gravitational waves exhibit a blue spectrum
and hence are relevant only at high frequencies.

\begin{figure}[tbp]
  \begin{center}
  \includegraphics[keepaspectratio=true,width=80mm]{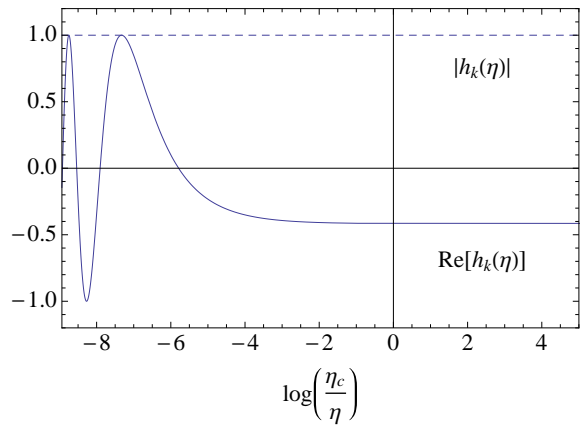}
  \end{center}
  \caption{The numerical
  evolution of gravitational waves during generalized Galilean genesis with $\alpha =1$.
  $\eta_c$ denotes the horizon-crossing time.}%
  \label{fig:mattercreation_h.eps}
\end{figure}

The quadratic action for the gravitational waves in generalized Galilean
genesis is~\cite{Nishi:2015pta}
\begin{eqnarray}
S_h^{(2)}=\frac{1}{8}\int \D t\D^3x \,a^3{\cal G}(Y_0)\left[\dot h^2_{ij}-\frac{c_t^2}{a^2}(\partial_k h_{ij})^2\right],
\label{action_hij}
\end{eqnarray}
where $c_t^2:=[\mpl^2+4\lambda Y_0g_5(Y_0)]/{\cal G}(Y_0)$.
Note that the coefficients ${\cal G}(Y_0)$ and $c_t^2$ are constant during the genesis phase.
The equation of motion for
each Fourier mode of two polarization states, $h_k^\lambda$ ($\lambda = +,\times$),
reads
%\begin{eqnarray}
%\ddot h_k + 3H\dot h_k + \frac{c_t^2k^2}{a^2}h_k = 0,
%\end{eqnarray}
\begin{align}
\frac{1}{a}\left(a h_k \right)''+\left(c_t^2k^2-\frac{a''}{a}\right)h_k=0,
\end{align}
where
\begin{align}
\frac{a''}{a}\simeq 
\frac{h_0}{(-a_G\eta)^{2\alpha}}\cdot\frac{2\alpha +1}{\eta^2}
\label{dda}
\end{align}
and $\lambda$ is omitted here and hereafter.
From Eq.~(\ref{dda}) it can be seen that the gravitational waves
freeze not at $|c_t k\eta|\sim 1$ but at later times,
$|c_t k\eta|\sim\sqrt{h_0/(-a_G\eta)^{2\alpha}}\ll 1$.
The WKB solution for the subhorizon modes is given by
\begin{eqnarray}
h_k=\frac{1}{a}\sqrt{\frac{2}{{\cal G}c_t k}}e^{-ic_t k\eta}.
\end{eqnarray}
Since $a\simeq a_G$, the power spectrum is already constant at early times,
\begin{eqnarray}
\frac{k^3}{\pi^2}|h_k|^2=\frac{2k^2}{\pi^2{\cal G} c_ta_G^2}.
\end{eqnarray}
After each mode exits the ``horizon'' and ceases to oscillate,
this amplitude is retained, as illustrated by the numerical example in Fig.~\ref{fig:mattercreation_h.eps}.
Thus, the power spectrum of the primordial gravitational waves
evaluated at the end of the genesis phase is given by
\begin{eqnarray}
{\cal P}_h^{(p)}(k)=\frac{2k^2}{\pi^2{\cal G} c_ta_G^2}.
\end{eqnarray}
In contrast to the gravitational waves from inflation,
the primordial spectrum is blue, and hence
gravitational waves could be relevant observationally only at high frequencies.

%----------------------------------------------------------------%
%----------------------------------------------------------------%

The density parameter for the gravitational waves
per log frequency interval, $\Omega_{\rm gw}$, evaluated
at the present epoch is given by
\begin{eqnarray}
\Omega_{\rm gw}=\frac{k^2}{12H_0^2}{\cal P}_h(k),
\end{eqnarray}
where $H_0$ is the present value of the Hubble parameter and
the power spectrum ${\cal P}_h$ characterizes
the present amplitude of the gravitational waves.
Since $h_k = \;$ const on superhorizon scales and
$h_k\propto a^{-1}$ after horizon re-entry,
we have
\begin{eqnarray}
{\cal P}_h(k)={\cal P}^{(p)}_h(k)\left(\frac{a_k}{a_0}\right)^2,
\end{eqnarray}
where $a_k$ is the scale factor at horizon re-entry
and $a_0\,(=1)$ is the scale factor at present.

\begin{figure}[tbp]
  \begin{center}
  \includegraphics[keepaspectratio=true,width=80mm]{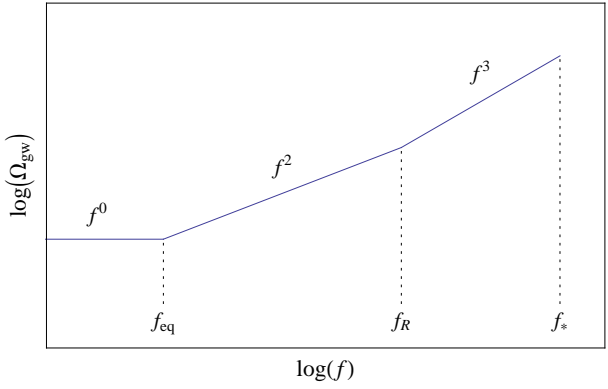}
  \end{center}
  \caption{Schematic shape of the power spectrum of gravitational waves in generalized Galilean genesis. }%
  \label{fig:mattercreation_omegagw.eps}
\end{figure}

Suppose that a mode with wavenumber $k$
re-enters the horizon at the epoch characterized by
the equation-of-state parameter $w$.
Since $k=a_k H_k$, where
the Hubble parameter at horizon re-entry obeys $H_k\propto a_k^{-3(1+w)/2}$,
we find
\begin{eqnarray}
a_k\propto k^{-2/(1+3w)}.
\end{eqnarray}
In our scinario we have the kination phase ($w=1$) after the end of genesis
and subsequently the conventional radiation and matter dominated phases.
Thus, we obtain
\begin{eqnarray}
\Omega_{\rm gw}=\Omega_{\rm gw}^{(p)}(k)\times \left \{
\begin{array}{ll}
\displaystyle
	\frac{k_R}{k}\frac{k_{\rm eq}^2}{k_R^2}\frac{k_0^4}{k_{\rm eq}^4} & (k_R < k < k_*)
 \\
\displaystyle
	\frac{k_{\rm eq}^2}{k^2}\frac{k_0^4}{k_{\rm eq}^4} & (k_{\rm eq} < k < k_R)
 \\
\displaystyle
	\frac{k_0^4}{k^4} & (k_0 < k < k_{\rm eq}),
\label{omegak_omgegap}
\end{array}
\right.
\end{eqnarray}
where we write
\begin{eqnarray}
\Omega_{\rm gw}^{(p)}(k)=\frac{k^2}{12H_0^2}{\cal P}_h^{(p)}=
\frac{k^4}{6\pi^2H_0^2{\cal G}c_t a_G^2}.
\end{eqnarray}
Here, the wavenumbers $k_\ast$, $k_R$, $k_{\rm eq}$, and $k_0$
correspond to the modes that re-enter the horizon
at the end of genesis, at the reheating time, at
the epoch of matter-radiation equality, and at the present epoch.
Explicitly, we have
\begin{align}
 k_0&= a_0H_0 =2.235\times 10^{-4}\left(\frac{h}{0.67}\right)\;{\rm Mpc}^{-1},\\
 k_{\rm eq}&=a_{\rm eq}H_{\rm eq}=1.028\times 10^{-2}\left(\frac{\Omega_{\rm m}h^2}{0.141}\right)\;{\rm Mpc}^{-1}.
\end{align}
It is also useful to write
\begin{align}
\frac{k_\ast}{k_R}&=\left(\frac{a_R}{a_G}\right)^2
\notag \\ &=
3^{\frac{1+\alpha}{2+\alpha}}\left(\frac{32\pi^2}{A}\right)^{\frac{1+2\alpha}{2(2+\alpha)}}
\left(\frac{\pi^2g_\ast}{30}\right)^{-\frac{1}{2(2+\alpha)}}\times
\notag \\ &\quad \times
\tilh^{\frac{1}{2+\alpha}}
\left(\frac{T_R}{\mpl}\right)^{-\frac{2}{2+\alpha}},\label{kast-kr}
\end{align}
where we used the fact that the Hubble parameter at the reheating time, $H_R$,
follows from $H_R = H_\ast(a_G/a_R)^3$.
In terms of
the frequency, $k_R$ corresponds to 
\begin{eqnarray}
f_R \simeq 0.026\left(\frac{g_*}{106.75}\right)^{1/6}\left(\frac{T_R}{10^6 \,{\rm GeV}}\right)\,{\rm Hz}. \label{fR_TR}
\end{eqnarray}
Using~(\ref{kast-kr}) one can also write $f_\ast$.
A schematic picture of $\Omega_{\rm gw}$
as a function of the frequency $f$ is shown in Fig.~\ref{fig:mattercreation_omegagw.eps}.
This should be compared with the standard prediction from inflation,
where one has a flat spectrum for $f_{\rm eq}<f<f_R$~\cite{Maggiore:1999vm}.

%----------------------------------------------------------------%
%----------------------------------------------------------------%

%As we discussed in the beginning of this section, the spectrum is very small for the low frequency.
%According to Fig.\ref{fig:mattercreation_omegagw.eps}, however, the spectrum grows rapidly in $f_R \ll f \ll f_*$. 
%Thus there is a possibility that we can detect the gravitational waves. 

To see the possibility of detecting the gravitational waves from
generalized Galilean genesis,
let us evaluate $\Omega_{\rm gw}$ at $k=k_R$ and $k=k_\ast$.
For simplicity, we assume that ${\cal G}= \mpl^2$ and $c_t= 1$.
This is exact for $g_5=0$ models, and extending the following estimate
to the $g_5\neq 0$ case is straightforward.
%Omega_k_R
Then, it can be seen from Eq.~(\ref{omegak_omgegap}) that
\begin{align}
\Omega_{\rm gw}(k_R)
%&=&\frac{k_R^4}{6\pi^2H_0^2M_{Pl}^2a_G^2}\frac{k_{eq}^2}{k_R^2}\frac{k_0^4}{k_{eq}^4}\nonumber \\
&=\frac{1}{6\pi^2\mpl^2H_0^2a_G^2}\frac{k_R^2k_0^4}{k_{eq}^2}
\notag \\
%&\simeq 10^{-5}\times  \left(\frac{H_R}{\mpl}\right)^2\left(\frac{a_R}{a_G}\right)^{2}
%\notag \\
&\simeq 10^{-5}\times  \left(\frac{H_\ast}{\mpl}\right)^2\left(\frac{a_R}{a_G}\right)^{-4},
\\
\Omega_{\rm gw}(k_\ast)
&=\Omega_{\rm gw}(k_R)\times \frac{k_\ast^3}{k_R^3}
\notag \\
%&\simeq 10^{-5}\times  \left(\frac{H_R}{\mpl}\right)^2\left(\frac{a_R}{a_G}\right)^{8}
%\notag \\
&\simeq 10^{-5}\times  \left(\frac{H_\ast}{\mpl}\right)^2\left(\frac{a_R}{a_G}\right)^{2},
\end{align}
where we used $k_0=H_0$.
The ratio $a_R/a_G\;(>1)$ itself depends on $H_\ast$ and model parameters through Eq.~(\ref{aragratio}).
At this stage
it is instructive to compare those results with the
flat spectrum in the inflationary scenario,
\begin{align}
\Omega_{\rm gw}^{\rm inf}\simeq 10^{-5}\times\left(\frac{H_{\rm inf}}{\mpl}\right)^2.
\end{align}
For fixed $T_R$, it follows from Eq.~(\ref{trh2}) that $H_\ast<H_{\rm inf}$.
This shows that $\Omega_{\rm gw}(k_R)$ in generalized Galilean genesis
is smaller than the inflationary prediction for fixed $T_R$.
However, this does not hold true for $\Omega_{\rm gw}(k_\ast)$.

Now, by the use of Eqs.~(\ref{aragratio}) and~(\ref{trh1}) we obtain
\begin{align}
\Omega_{\rm gw}(k_R)&=10^{-5}\cdot
3^{-\frac{1}{2+\alpha}}\left(\frac{32\pi^2}{A}\right)^{\frac{1+2\alpha}{2(2+\alpha)}}
\left(\frac{\pi^2g_\ast}{30}\right)^{\frac{3+2\alpha}{2(2+\alpha)}}
\notag \\ &\quad\times
\tilh^{\frac{1}{2+\alpha}}
\left(\frac{T_R}{\mpl}\right)^{\frac{2(3+2\alpha)}{2+\alpha}},
\end{align}
and
\begin{align}
\Omega_{\rm gw}(k_\ast)&=10^{-5}\cdot
3^{\frac{2+3\alpha}{2+\alpha}}\left(\frac{32\pi^2}{A}\right)^{\frac{2(1+2\alpha)}{2+\alpha}}
\left(\frac{\pi^2g_\ast}{30}\right)^{\frac{\alpha}{2+\alpha}}
\notag \\ &\quad\times
\tilh^{\frac{4}{2+\alpha}}
\left(\frac{T_R}{\mpl}\right)^{\frac{4\alpha}{2+\alpha}}.
\end{align}
In the frequency range $f_R<f<f_*$, we have
\begin{align}
\Omega_{\rm gw}(f)&=10^{-31}\cdot
3^{-\frac{1}{2+\alpha}}\left(\frac{32\pi^2}{A}\right)^{\frac{1+2\alpha}{2(2+\alpha)}}
\left(\frac{\pi^2g_\ast}{30}\right)^{\frac{1+\alpha}{2(2+\alpha)}}
\notag \\ &\quad\times
\tilh^{\frac{1}{2+\alpha}}
\left(\frac{T_R}{\mpl}\right)^{\frac{\alpha}{2+\alpha}}\left(\frac{f}{100\,{\rm Hz}}\right)^3.
\end{align}
%%%

One could enhance $\Omega_{\rm gw}$ by taking large $\tilh$,
but too large $\tilh$ would violate Eq.~(\ref{gen-an-cond2}),
indicating the breaking of the approximation $a\simeq a_G[1+h_0/2\alpha (-t)^{2\alpha}]$.
Using Eqs.~(\ref{gen-an-cond2}) and~(\ref{trh1}) we have, for $f_R<f<f_\ast$,
\begin{align}
\Omega_{\rm gw}(f)
\lesssim 10^{-30}A^{-1/4}\left(\frac{g_\ast}{106.75}\right)^{1/4}\left(\frac{f}{100\,{\rm Hz}}\right)^3.
\label{bound1}
\end{align}
Note that the right hand side is independent of $\alpha$ and the reheating temperature.
To generate a higher amplitude of the gravitational waves,
one would relax the limitation imposed by Eq.~(\ref{gen-an-cond2}).

In the above argument we have assumed $c_t=1$ for simplicity, 
but it is worth noting that we can enhance the amplitude of the gravitational waves by assuming that $c_t<1$.

%--- Examples ---%

\section{Examples}

To be more specific, let us move to the discussion
of concrete examples in the this section.

\subsection{The original model of Galilean genesis}
First, we apply the above results to the original model of Galilean genesis~\cite{Creminelli:2010ba},
which corresponds to the $\alpha =1$ case. 
The original model is given by choosing the arbitrary functions and parameters in (\ref{gen:Lag}) as
\begin{eqnarray}
&& g_2=-2\mu^2Y+\frac{2\mu^3}{\Lambda^3}Y^2,\quad g_3=\frac{2\mu^3}{\Lambda^3}Y, \nonumber \\ 
&& g_4=g_5=0,\quad \lambda=1, \quad\alpha =1,
\end{eqnarray}
where $\mu$ and $\Lambda$ are the parameters having the dimension of mass.
Note that $\phi$ is taken to be dimensionless and so $Y=\;$[mass]$^2$.
It is easy to verify that the model satisfies the stability conditions for scalar and tensor perturbations.
From Eqs.~(\ref{gen:eq1}) and~(\ref{h0determined}) we find 
\begin{eqnarray}
Y_0=\frac{\Lambda^3}{3\mu},\quad h_0=\frac{\mu^3}{2\mpl^2\Lambda^3}.
\end{eqnarray}
%We thus have
%\begin{eqnarray}
%B=\frac{1}{2}\left(\frac{\mu}{\Lambda}\right)^3.
%\end{eqnarray}
%In the $\alpha = 1$ model
%the final result depends on the parameters $\mu$ and $\Lambda$
%only through the ratio $\mu/\Lambda$.

Assuming that $A={\cal O}(1)$, the reheating temperature can be written as
\begin{eqnarray}
T_R\simeq 10^{-3} \left(\frac{g_\ast}{106.75}\right)^{-1/4}\left(\frac{\Lambda}{\mu}\right)^{3/2}H_\ast.
\end{eqnarray}
The condition that the analytic approximation for the genesis background solution is valid,
Eq.~(\ref{gen-an-cond2}), translates to
\begin{eqnarray}
\frac{\mu}{\Lambda} < 10^2 \left(\frac{T_R}{10^{10}{\rm GeV}}\right)^{-1/3}
\left(\frac{g_*}{106.75}\right)^{-1/12},\label{const1}
\end{eqnarray}
while $a_R/a_G>1$ yields
\begin{eqnarray}
\frac{\mu}{\Lambda} > 10^{-7}
 \left(\frac{T_R}{10^{10}{\rm GeV}}\right)^{2/3}\left(\frac{g_*}{106.75}\right)^{1/6}.
\end{eqnarray}
The density parameters at $k=k_R$ and $k= k_\ast$ are given respectively by
\begin{align}
\Omega_{\rm gw}(k_R)\simeq 10^{-31} 
\left(\frac{g_\ast}{106.75}\right)^{5/6}
\left(\frac{\mu}{\Lambda}\right)\left(\frac{T_R}{10^{10}\,{\rm GeV}}\right)^{10/3},
\end{align}
and
\begin{align}
\Omega_{\rm gw}(k_\ast)\simeq 10^{-10} 
\left(\frac{g_\ast}{106.75}\right)^{1/3}
\left(\frac{\mu}{\Lambda}\right)^4\left(\frac{T_R}{10^{10}\,{\rm GeV}}\right)^{4/3},
\end{align}
where the corresponding frequency is
\begin{align}
f_\ast=10^9\left(\frac{\mu}{\Lambda}\right)\left(\frac{T_R}{10^{10}\,{\rm GeV}}\right)^{1/3}{\rm Hz}.
\end{align}
In the frequency range $f_R<f<f_*$ we have
\begin{align}
\Omega_{\rm gw}(f)&\simeq 10^{-32}
\left(\frac{g_\ast}{106.75}\right)^{1/3}\left(\frac{\mu}{\Lambda}\right) \nonumber \\
&\quad
\times \left(\frac{T_R}{10^{10}\,{\rm GeV}}\right)^{1/3}\left(\frac{f}{100\,{\rm Hz}}\right)^3.
\label{omf1mod}
\end{align}

%----------------------------------------------------------------%
\begin{figure}[tbp]
  \begin{center}
  \includegraphics[keepaspectratio=true,width=85mm]{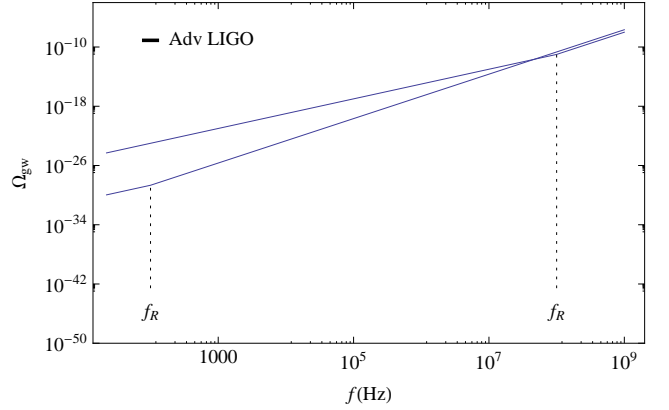}
  \end{center}
  \caption{The density parameters $\Omega_{\rm gw}(f)$ for different reheating temperatures in
  the original model of Galilean genesis. The two plots correspond to
  $T_R\sim 10^{10}$\;GeV ($f_R=100$\;Hz) with $\mu/\Lambda=10^2$
  and
  $T_R\sim 10^{16}$\;GeV ($f_R=10^8$\;Hz) with $\mu/\Lambda=1$.
  Here $\mu/\Lambda$ is taken to be the largest value allowed by Eq.~(\ref{const1}).
  The anticipated sensitivity of the advanced LIGO is also marked. }%
  \label{fig:omegagw_ex1.eps}
\end{figure}
%----------------------------------------------------------------%

Figure~\ref{fig:omegagw_ex1.eps}
shows the examples of the power spectra compared with the anticipated sensitivity of the advanced LIGO.
We consider two different reheating temperatures.
One is $T_R\sim 10^{10}$\;GeV and it follows from Eq.~(\ref{fR_TR}) that
the corresponding frequency is given by $f_R=100$\;Hz.
In this case, the amplitude is too small to be detected by the advanced LIGO.
The other is as high as $T_R\sim 10^{16}$\; GeV, giving $f_R=100$\;MHz.
Also in this case it is unlikely to be able to detect the primordial gravitational waves
at $f=100$\;Hz.
In both cases
we have $\Omega_{\rm gw}\sim 10^{-12}$ at $f=100$\;MHz,
as can be seen directly from Eq.~(\ref{bound1}).
The gravitational reheating stage after
Galilean genesis could therefore be probed by
the gravitational waves at very high frequencies.

%This difference of $f_R$ means the difference of the reheating temperature $T_R$ as Eq.\ref{fR_TR}.
%We set the frequency $f_R$ as $f_R=100{\rm Hz}$ or $f_R=10^7 {\rm Hz}$.
%From fig.\ref{fig:omegagw_ex1.eps}, We can find it is difficult to observe this by present detectors.
%However, at $f\simeq 10^8$ the spectrum grows approximately $\Omega_{gw}\simeq 10^{-10}$. 
%Thus, if we have the detector that can observe the high frequency area, we can find this model.

\subsection{A model generating the scale invariant curvature perturbation}% 2nd Example

As a second example, let us consider the case of $\alpha=2$.
This class of models is of particular interest because it
gives rise to a scale-invariant spectrum of the curvature perturbation~\cite{Liu:2011ns,Nishi:2015pta}.
The Lagrangian is given in a similar form to the previous example by
\begin{eqnarray}
&& g_2=-2\mu^2Y+\frac{2\mu^3}{\Lambda^3}Y^2,\quad g_3=\frac{\mu^3}{\Lambda^3}Y, \nonumber \\ 
&& g_4=g_5=0,\quad \lambda=1,
\end{eqnarray}
but now $\alpha = 2$.
It follows from Eqs.~(\ref{gen:eq1}) and~(\ref{h0determined}) that
\begin{eqnarray}
Y_0=\frac{\Lambda^3}{\mu},\quad h_0=\frac{\mu^4}{10\mpl^2\Lambda^6}.
\end{eqnarray}
%and so we may take
%\begin{eqnarray}
%B=\frac{1}{10}\left(\frac{\mu}{\Lambda}\right)^6.
%\end{eqnarray}
%One can verify that the stability conditions for scalar and tensor perturbations are satisfied.

The reheating temperature is obtained as
\begin{align}
 T_R\simeq10^{-2}\left(\frac{g_\ast}{106.75}\right)^{-1/4}\frac{\Lambda^{9/5}H_\ast^{4/5}}{\mpl^{2/5}\mu^{6/5}}.
\end{align}
Equation~(\ref{gen-an-cond2}) in this case
reduces to 
\begin{eqnarray}
\frac{\mpl \mu^2}{\Lambda^3} < 10^6 \left(\frac{T_R}{10^{10}{\rm GeV}}\right)^{-1}\left(\frac{g_*}{106.75}\right)^{-1/4},\label{co_2_1}
\end{eqnarray}
while
$a_R/a_G>1$ gives 
\begin{eqnarray}
\frac{\mpl \mu^2}{\Lambda^3} > 10^{-11} \left(\frac{T_R}{10^{10}{\rm GeV}}\right)\left(\frac{g_*}{106.75}\right)^{1/4}.\label{co_2_2}
\end{eqnarray}
The density parameters at $k=k_R$ and $k=k_\ast$ are computed respectively as
\begin{align}
\Omega_{\rm gw}(k_R)\simeq 10^{-32} 
\left(\frac{g_\ast}{106.75}\right)^{7/8}
\left(\frac{\mpl \mu^2}{\Lambda^3}\right)^{1/2}\left(\frac{T_R}{10^{10}\,{\rm GeV}}\right)^{7/2},
\end{align}
and
\begin{align}
\Omega_{\rm gw}(k_\ast)\simeq 10^{-15} 
\left(\frac{g_\ast}{106.75}\right)^{1/2}
\left(\frac{\mpl \mu^2}{\Lambda^3}\right)^2\left(\frac{T_R}{10^{10}\,{\rm GeV}}\right)^{2},
\end{align}
with the corresponding frequency
\begin{align}
f_\ast=10^{8}\left(\frac{g_\ast}{106.75}\right)^{1/24}
\left(\frac{\mpl \mu^2}{\Lambda^3}\right)^{1/2}\left(\frac{T_R}{10^{10}\,{\rm GeV}}\right)^{1/2}\,{\rm Hz}.
\end{align}
In the range $f_R<f<f_*$, we find
\begin{align}
\Omega_{\rm gw}(f)&\simeq 10^{-33} \left(\frac{g_\ast}{106.75}\right)^{3/8}\left(\frac{\mpl \mu^2}{\Lambda^3}\right)^{1/2} \quad\notag \\ &\quad
\times \left(\frac{T_R}{10^{10}\,{\rm GeV}}\right)^{1/2}\left(\frac{f}{100\,{\rm Hz}}\right)^3.
\end{align}

%----------------------------------------------------------------%
\begin{figure}[tbp]
  \begin{center}
  \includegraphics[keepaspectratio=true,width=85mm]{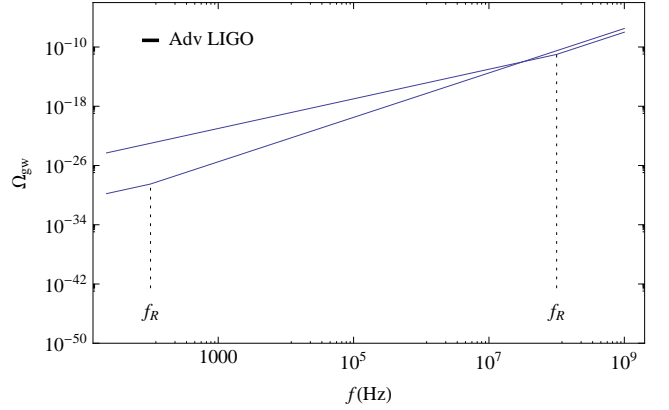}
  \end{center}
  \caption{The density parameters $\Omega_{\rm gw}(f)$
  for different reheating temperatures in the $\alpha=2$ model,
  in comparison with the anticipated sensitivity of the advanced LIGO.
  The parameter $\mpl \mu^2/\Lambda^3$ is taken to be the possible largest values:
  $\mpl \mu^2/\Lambda^3=10^6$ for $T_R\sim 10^{10}$\;GeV ($f_R=100$\;Hz)
  and
  $\mpl \mu^2/\Lambda^3=1$ for $T_R\sim 10^{16}$\;GeV ($f_R=10^8$\;Hz).
  }%
  \label{fig:omegagw_ex2.eps}
\end{figure}
%----------------------------------------------------------------%

The basic conclusions for the $\alpha=2$ model is the same as those for $\alpha=1$,
as presented in Fig.~\ref{fig:omegagw_ex2.eps}.
Though detectable gravitational waves are not expected
in the sensitive bands of, say, the LIGO detector, $\Omega_{\rm gw}$
is enhanced at high frequencies up to
the model independent value as shown in Eq.~(\ref{bound1}) and
$\Omega_{\rm gw}\sim 10^{-12}$ at $f=100$\;MHz
under the optimistic choice of the parameters.

%We set the frequency $f_R$ as $f_R=100{\rm Hz}$ or $f_R=10^7 {\rm Hz}$.
%From fig.\ref{fig:omegagw_ex2.eps}, We can find it is difficult to observe this by present detectors.
%However, at $f\simeq 10^8$ the spectrum grows approximately $\Omega_{gw}\simeq 10^{-10}$. 
%Thus, if we have the detector that can observe the high frequency area, we can find this model.

% shows the concrete examples in comparison with the sensitivity of advanced LIGO.

Since the $\alpha = 2$ model can generate scale-invariant curvature perturbations,
let us now consider consistency between the above optimistic estimate of $\Omega_{\rm gw}$
and the prediction for the curvature perturbations.
The power spectrum of the curvature perturbation $\zeta$
evaluated at the end of Galilean genesis is given by~\cite{Nishi:2015pta}
\begin{align}
\left.{\cal P}_\zeta\right|_{t=t_*}
=\frac{\Gamma^2(-3/2)}{2\pi^3c_s^3{\cal A}}\left(-t_*\right)^{-6},
\end{align}
where
in the present model it is found that $c_s=\sqrt{3}$ and ${\cal A}=(100/9)(\mpl\mu^2/\Lambda^3)^2$.
For simplicity we assume that the evolution of $\zeta$ in the post-genesis phase is negligible.
We can write $t_\ast$ in terms of $T_R$ to obtain
\begin{align}
{\cal P}_\zeta\simeq 10^{-11}\left(\frac{g_*}{106.75}\right)^{3/8}
\left(\frac{\mpl \mu^2}{\Lambda^3}\right)^{1/2}
\left(\frac{T_R}{10^{10}\,{\rm GeV}}\right)^{3/2}.
\end{align}
Using ${\cal P}_\zeta \sim 10^{-9}$, the parameters are fixed as
\begin{eqnarray}
\frac{\mpl \mu^2}{\Lambda^3}\sim 10^4\left(\frac{T_R}{10^{10}\,{\rm GeV}}\right)^{-3}.
\end{eqnarray}
This gives smaller values of $\mpl \mu^2/\Lambda^3$ than the most optimistic ones
used in Fig.~\ref{fig:omegagw_ex2.eps}. For $T_R\sim 10^{10}$\,GeV we have $\Omega_{\rm gw}\sim 10^{-13}$
at $f=100$\,MHz.
If one would use the curvaton mechanism to produce the observed spectrum of the curvature perturbation
even in the case of $\alpha =2$, one has ${\cal P}_\zeta<10^{-9}$ but then
$\mpl \mu^2/\Lambda^3< 10^4(T_R/100\,{\rm GeV})^{-3}$.

%--- Conclusions ---%
\section{Conclusions}
In this paper, we have studied the reheating stage after the initial quasi-Minkowski expanding phase
and its consequences on the primordial gravitational waves.
The initial stage was described in a unified manner as {\em generalized Galilean genesis}~\cite{Nishi:2015pta}
in terms of the Horndeski scalar-tensor theory.
Since the scalar field does not have a potential in which it oscillates,
we have considered reheating through
the gravitational production of massless scalar particles
at the transition from the genesis phase to the kination phase.
To avoid an unphysical diverging result which would come from a sudden transition approximation,
we have followed the previous works~\cite{Ford:1986sy,Kunimitsu:2012xx} and considered a smooth matching of
the two phases. We have computed in that way the energy density of created particles
and the reheating temperature.
In the case of gravitational reheating after inflation,
it is known that the created energy density is given by $\sim H_{\rm inf}^4$,
where $H_{\rm inf}$ is the inflationary Hubble parameter~\cite{Ford:1986sy}.
We have shown that the energy density of massless scalar particles
created after Galilean genesis is not simply given by the naive replacement,
$\sim H_\ast^4$, where $H_\ast$ is the Hubble parameter at the end of the genesis phase,
but rather by a more involved form which depends on the model parameters as well as $H_\ast$.
In particular, it has been found
that for the same reheating temperature
the Hubble parameter at the end of genesis is smaller compared with
the corresponding value in the inflation scenario.

We have then discussed the spectrum of the primordial gravitational waves
from generalized Galilean genesis.
The combined effects of the quasi-Minkowski expanding background and the kination phase
give rise to the blue gravitational waves, $\Omega_{\rm gw}\propto f^3$,
at high frequencies, while their amplitude is highly suppressed at low frequencies
in contrast with the inflationary gravitational waves.
Unfortunately, the expected amplitude is too small to be detected
in the sensitive bands of the advanced LIGO detector. However,
it is possible to have $\Omega_{\rm gw} \sim 10^{-12}$ at $f\sim 100$\,MHz.
Thus, the primordial gravitational waves having a spectrum $\Omega_{\rm gw}\propto f^3$
at $f\gtrsim 100$\,MHz
and the lack thereof at low frequencies
would offer an interesting test of Galilean genesis in future experiments.

Finally, let us comment that the equations of motion for a massless scalar field and
gravitational waves are practically the same, though we have discussed
the particle production of the former on subhorizon scales while the latter cross the horizon.
This implies that the gravitons can also be generated on subhorizon scales at the transition between the two phases
in the same way as the massless scalar field,
and this concern must be taken care of in the gravitational reheating scenario.

%--- Acknowledgements ---%--- Acknowledgements ---%--- Acknowledgements ---%
\acknowledgments
We thank Kazunori Nakayama, Teruaki Suyama, Masahide Yamaguchi, Jun'ichi Yokoyama, and
Shuichiro Yokoyama for helpful conversations.
This work was supported in part by the JSPS Research Fellowships for Young Scientists No.~15J04044 (S.N.)
and the JSPS Grant-in-Aid for Young Scientists (B) No.~24740161 (T.K.).
% 
%--- Acknowledgements ---%--- Acknowledgements ---%--- Acknowledgements ---%

%-------------------------------------------------------------------%

%\appendix

%-------------------------------------------------------------------%

%---------   References   ---------%

\end{document}